\documentclass[journal ]{new-aiaa}
\usepackage[utf8]{inputenc}
\usepackage{textcomp}
\usepackage{subfig}
\usepackage{caption}
\usepackage{graphicx}
\usepackage{amsmath}
\usepackage[version=4]{mhchem}
\usepackage{siunitx}
\usepackage{longtable,tabularx}
\usepackage{float}
\usepackage{multirow}
\usepackage{tabularray}
\usepackage{dcolumn,eurosym,rotating,booktabs}
\allowdisplaybreaks
\usepackage{cleveref}
\crefname{equation}{Eq}{Eqs} 
\crefrangelabelformat{equation}{(#3#1#4--#5#2#6)}
\usepackage[LGR]{mathastext}
\usepackage{footnotehyper}

\title{Effect of non-equilibrium thermochemistry \\
on Pitot pressure measurements in shock tunnels \\
(or: Is 0.92 really the magic number?)}

\author{Tamara Sopek \footnote{Postdoctoral Research Fellow, School of Engineering.},
Peter Jacobs \footnote{Assoc. Professor, Institute for Advanced Engineering and Space Sciences}}
\affil{University of Southern Queensland, Toowoomba, Queensland, 4350, Australia}

\author{Suria-Devi Subiah\footnote{DPhil student, Department of Engineering Science}, Peter Collen\footnote{Postdoctoral Research Assistant},
	and Matthew McGilvray\footnote{Professor, Department of Engineering Science.}}
\affil{University of Oxford, Oxford, OX1 2JD, UK}

\begin{document}

\maketitle

\begin{abstract}
\noindent Pitot pressure is the most common measurement in high total enthalpy shock tunnels for test condition verification. Nozzle calculations using multi-temperature non-equilibrium thermochemistry are needed in conjunction with Pitot measurements to quantify freestream properties. Pitot pressure is typically matched by tuning the boundary layer transition location in these simulations. However, non-equilibrium thermochemistry effects on the Pitot probe are commonly ignored. A computational study was undertaken to estimate the effect of non-equilibrium thermochemistry on Pitot pressure and freestream conditions. The test flow was produced by a Mach 7 nozzle in a reflected shock tunnel for air at a relatively low total enthalpy of 2.67\,MJ/kg. Three different thermochemical models (equilibrium, finite-rate chemistry and two-temperature thermochemistry) were employed to compute flow variables at the nozzle exit and Pitot probe. Pitot pressures from these simulations were compared against those obtained via experiments. The results show a departure from the commonly utilized C of 0.92 in the reduced Rayleigh-Pitot equation form for high Mach numbers. Additionally, calculations were done with a sweep of free-stream conditions and resulting in C$\approx$0.94. These results show that the influence of finite-rate thermochemistry should be taken into account, even at relatively low flow enthalpies. 
\end{abstract}

\section*{Nomenclature}

{\renewcommand\arraystretch{1.0}
\noindent\begin{longtable*}{@{}l @{\quad=\quad} l@{}}
$C_{p,\,max}$ & maximum value of the pressure coefficient \\
$C$ & constant, $C$=0.5*$C_{p,\,max}$ \\
$M$ & Mach number \\
$p$ & pressure, Pa \\
$T$ & temperature, K \\
$U$ & axial velocity, m/s \\
$\rho$ & density\\
\multicolumn{2}{@{}l}{Subscripts}\\
$tr$ & translation and rotational\\
$ve$ & vibrational and electronic\\
\multicolumn{2}{@{}l}{Abbreviations}\\
ESTCN & Equilibrium Shock Tube Conditions, with Nozzle\\
RST & reflected shock tunnel\\
NENZF1d & Non-Equilibrium Nozzle Flow, one dimension\\
\end{longtable*}}

\section{Introduction}

Shock tunnels are a class of impulse facilities used to investigate hypersonic flows \cite{Morgan2001a}. They present a fundamental instrument to replicate conditions encountered during high-speed flight. Because these facilities can simulate a range of temperatures and pressures and use different test gas compositions, they are a useful device for the investigation of conditions that might be difficult to achieve in other facilities. However, the flow state and variables must be determined to a sufficiently high level of accuracy to allow analysis of the obtained data with reasonably low uncertainty. Typically, Pitot pressure measurements are performed to verify the condition and size of the core flow at the nozzle exit. To estimate additional properties of the test flow, a computational study is usually undertaken from stagnation conditions as it is unpractical to measure all the relevant flow variables. However, the computed variables are highly dependent on the type of the gas model and boundary layer modeling. In hypersonic high temperature gases, the thermochemistry changes substantially with the increase of flow speed. These gas effects, such as excitation of vibrational energy modes, may affect the assumed Pitot pressure, though this is typically ignored.

The Rayleigh-Pitot formula \cite{NACA1135} shown in (Eq.~\ref{eq:Rayleigh-Pitot}) gives a relationship between stagnation and static pressure as a function of Mach number and ratio of specific heats. It is derived as the product of the normal shock and isentropic relations for pressure at stagnation point behind the normal shock wave, and, for an ideal gas can be written as: 
\begin{equation}
\frac{\textit{p}_{0,2}}{\textit{p}_1} = \left[ \frac{(1+\gamma)M_1^2} {2} \right]^{\frac{\gamma}{\gamma-1}} \left[ \frac{\gamma+1}{2 \gamma M^2_1-(\gamma-1)}\right]^{\frac{1}{\gamma-1}}
	\label{eq:Rayleigh-Pitot}
\end{equation} 

\noindent which, in the limit of high Mach number can be reduced to:
\begin{equation}
 \textit{p}_{0,2} = \textit{p}_{Pitot} = \frac{1}{2} \rho \textit{U}^2 \cdot C_{p,\,max} = C \rho \textit{U}^2
	\label{eq:Pitot}
\end{equation} 

\noindent where \begin{math}C_{p\,max}\end{math} and \begin{math}C=\frac{1}{2}C_{p,\,max}\end{math} are constants. \begin{math}C_{p,\,max}\end{math} is the maximum value of the pressure coefficient, evaluated at the stagnation point behind a normal shock wave. Full derivation of the maximum pressure coefficient, \begin{math}C_{p,\,max}\end{math}, is presented in the Appendix, in \cref{Cpmax,Cpmax-supersonic,Cpmax-infinite-Mach,Cpmax-infinite-Mach-2,Cpmax-infinite-Mach-3,Cpmax-infinite-Mach-4}.

It is shown in Eq.\ref{Cpmax-infinite-Mach-4} from the Appendix that in the limit of high hypersonic flows \begin{math}C_{p,\,max}\end{math} reaches the asymptotic value \begin{math}C_{p,\,max}=1.8394\end{math} for \begin{math}\gamma=1.4\end{math} \cite{Anderson2006}. 
C which is used to compute $\textit{p}_{Pitot}$ is then obtained by halving that value, i.e. \begin{math}$C$\,=0.9197\end{math} or more commonly a rounded value of 0.92 \cite{Fuller1961}. This value of 0.92 is for an ideal gas model, which implies that this value should be validated when the effect of thermochemistry present in real shock tunnel nozzles is in play. The practice of using the reduced form Rayleigh-Pitot equation goes back decades as seen in the tables starting at page 25 in Ref.\cite{Fuller1961}.

Kantrowitz \cite{Kantrowitz1946} conducted a study investigating the influence of non-equilibrium thermochemistry on energy dissipation. He performed measurements with CO$_2$ where the gas was expanded from a chamber through a nozzle, and compressed at the nose of a Pitot tube aligned with the jet from the nozzle. Pitot pressure was measured over a range of chamber pressures. The results show dependence of the resultant Pitot pressure on the relaxation time of the gas when the gas is in non-equilibrium, the measured Pitot pressure will be lower than the chamber pressure, and this ``loss'' of Pitot pressure increases with increased chamber pressure. These initial findings were confirmed by a subsequent study \cite{Huber1947} conducted with a range of gases. 

Work was done previously \cite{Wyllie2013} to investigate the effects of hypersonic and high temperature gas on Pitot pressure. A parametric study was completed by altering the flow conditions to isolate the effects of individual flow characteristics, for example by running simulations with thermal effects on, followed by thermal effects off. Calculations were performed for a range of conditions and two nozzles (M6 and M8) using both an inviscid MATLAB code and US3D code \cite{Candler2015}. The full-form for an ideal gas (Eq.~\ref{eq:Rayleigh-Pitot}) and reduced form (Eq.~\ref{eq:Pitot}) of Rayleigh-Pitot equation were used to assess their accuracy in predicting Pitot pressure without numerical modeling of Pitot probe. In that study it was suggested that a value of 0.94 be used for C in the reduced form of the Rayleigh-Pitot equation.

The aim of this work was to investigate the effect of non-equilibrium thermochemistry on Pitot pressure measurements in shock tunnels, as well as the validity of the Rayleigh-Pitot equation in both general and reduced form used in hypersonic flows for different thermochemical models. Numerical calculations were performed and compared against experimental data for a low enthalpy air condition in the Oxford T6 Stalker facility operated in RST mode. The nozzle-exit flow values were then subsequently used as an inflow for Pitot probe simulations. The reduced form Rayleigh-Pitot equation ($\textit{p}_{Pitot}$=C$\rho \textit{U}^2$) for hypersonic flows was calculated from the results of the simulations performed using three thermochemical models. 

\section{Numerical methodology}

Numerical results were obtained with two types of calculations: the first calculation was performed for just the nozzle, and then the second calculation was done using outflow from the nozzle simulation as an inflow for the Pitot probe simulation. Positioning of the nozzle and the Pitot probe was done to emulate that in measurements, as shown in Fig.~\ref{fig:grid}.

\begin{figure}[htbp]
	\centering
	\includegraphics[width=0.97\textwidth]{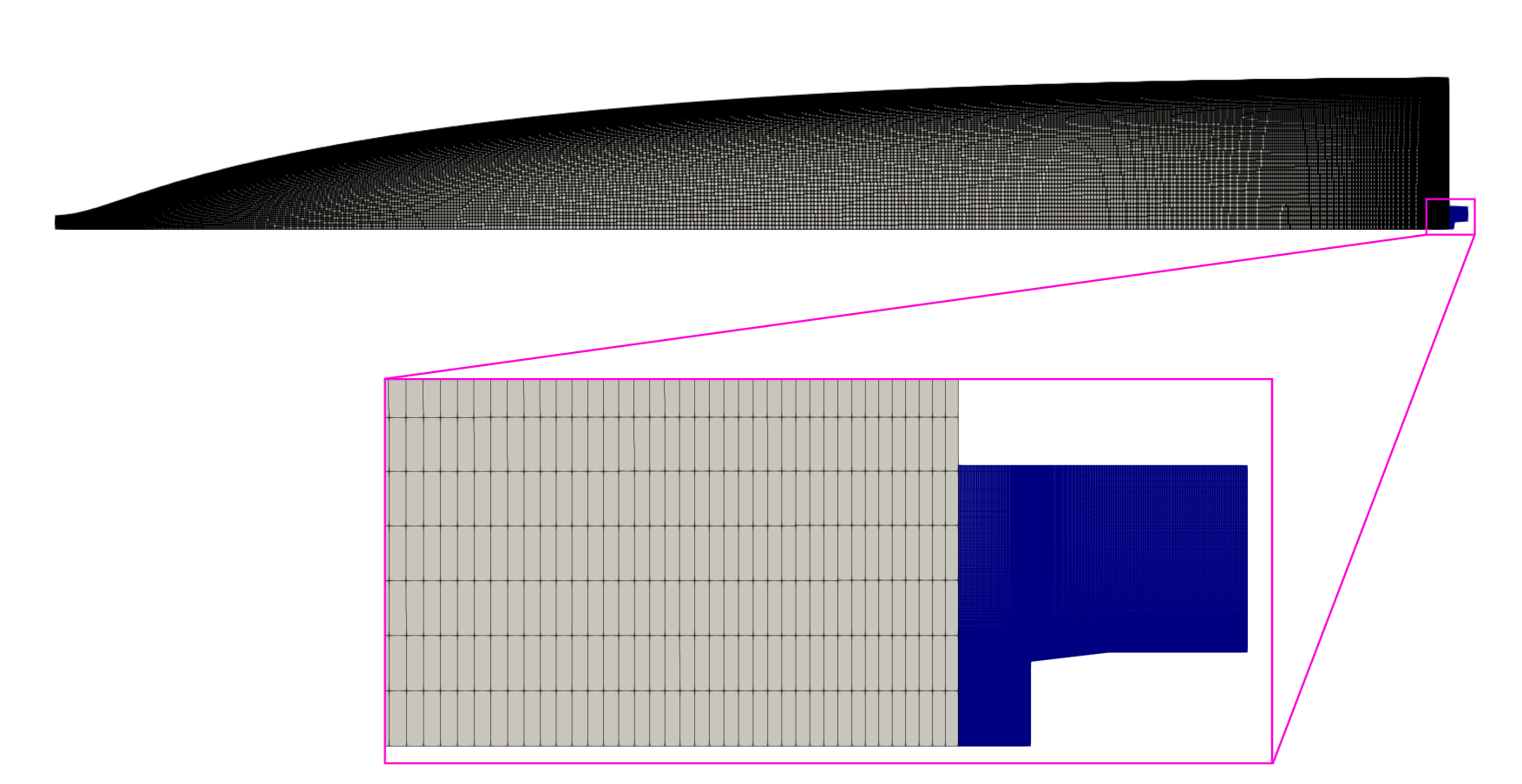}\label{fig:nozzle-and-probe}
	\caption{Placement of the nozzle and Pitot probe.}
	\label{fig:nozzle-and-probe}
\end{figure}

\subsection{Eilmer Code}
The computational study was undertaken using the Eilmer code \cite{Gibbons2022}, which is an open-source \footnote{Eilmer code is a part of the Gas Dynamics Toolkit and freely available at \url{https://gdtk.uqcloud.net}} Reynolds-averaged Navier-Stokes computational fluid dynamics solver. This CFD solver is a collection of codes for the numerical simulation of transient (time-accurate), compressible gas flows in two and three dimensions. It has been validated on numerous high-speed (supersonic and hypersonic) test cases, and it can be used to model these highly energetic flows in chemical and thermal equilibrium and non-equilibrium. This ability makes it ideal for high total enthalpy flows where the flow behind the shock wave experiences very high temperatures (sometimes in excess of 10000\,K) where non-equilibrium thermochemistry should be considered. 

\subsection{Thermochemistry modeling}
The simulations were performed with three different thermochemical gas models to investigate the influence of high-enthalpy chemical kinetics. These models include equilibrium thermochemistry (based on 5-species air model in the CEA2 program \cite{Gordon1994}), finite-rate chemistry (based on Gupta's 5-species air model \cite{Gupta1990}) and two-temperature thermochemistry (based on Park's 2-temperature 5-species air model \cite{Gupta1990}). These thermochemistry models are referred to, respectively, in this section of the paper as the equilibrium, chemical non-equilibrium and thermochemical non-equilibrium models. The mixture transport properties, that is, mixture viscosity and thermal conductivity, were calculated using mixing rules of Gordon and McBride \cite{Gordon1994} which are a variant of Wilke's original formulation \cite{Wilke1950}.

\subsection{Nozzle setup}\label{s:nozzle-setup}

\begin{figure}[htbp]
	\centering
	\subfloat[]{\includegraphics[width=0.97\textwidth]{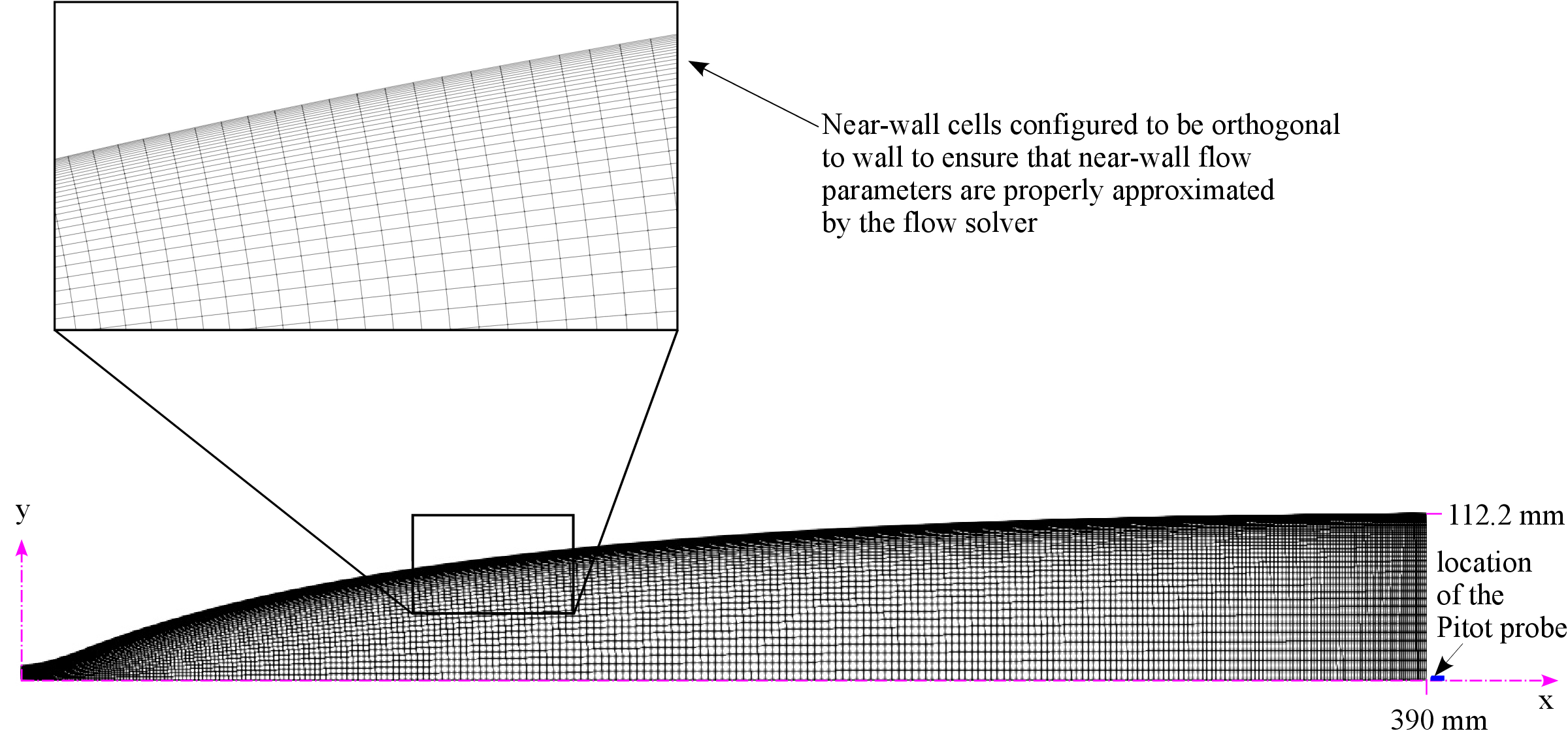}\label{fig:nozzle-grid}}
	\hfill
	\subfloat[]{\includegraphics[width=\textwidth]{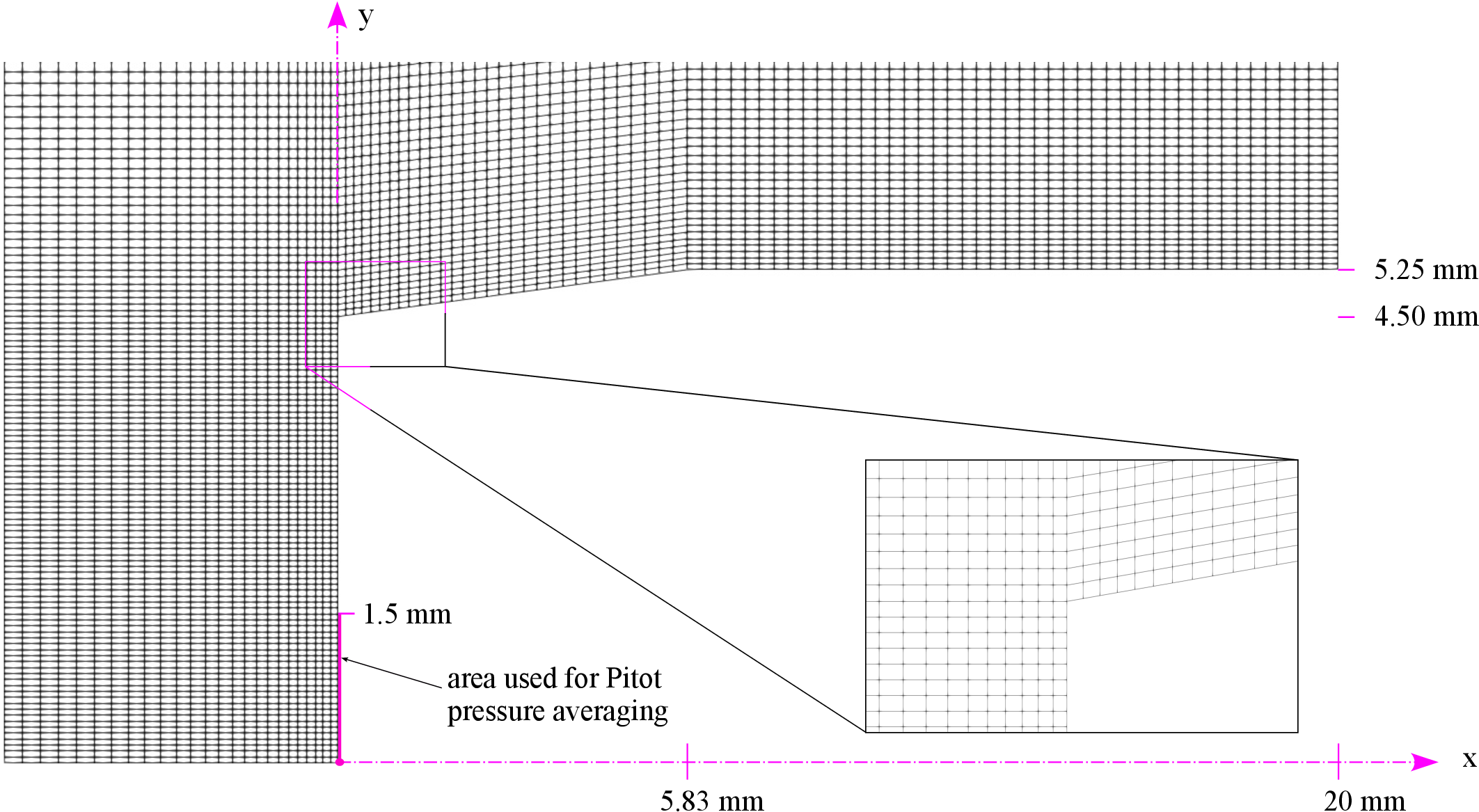}\label{fig:Pitot-grid}}
	\caption{Computational grid for a) Mach 7 nozzle and b) Pitot probe simulations. Inflow is from the left-hand boundary.}
	\label{fig:grid}
\end{figure}

For the axisymmetric nozzle simulations, the approach taken was identical to that used in previous studies of similar facilities, e.g. \cite{Chan2018}, where the nozzle is simulated from the throat onwards. It has been shown that the differences when including the region upstream of the nozzle vs not including it are very small, while the cost is significantly reduced when simulating from the throat onwards. The stagnation properties (Table~\ref{tab:stagnation-conditions}) are isentropically expanded to a Mach number of 1 to give a uniform inflow at the throat and that process is assumed to be in thermochemical equilibrium due to subsonic and relatively slow transient to reaction times. The wall of the nozzle was assumed to be at a constant temperature of 300\,K. Shock tunnels have short flow durations and the nozzle wall does not change in temperature by more than a few degrees during the test flow time. A pseudo-space-marching approach was used to reduce the computational time needed to obtain a steady-state nozzle flow solution \cite{Jacobs2020}. The turbulence model used in the simulations was the $k-\omega$ turbulence model of Wilcox \cite{Wilcox2006}. To further reduce computational time, the wall functions of Nichols \cite{Nichols2004} were used to model the turbulent boundary layers. The application of wall functions is appropriate since boundary layer separation is not expected in the nozzle flow field. The turbulence intensity of the inflow to the nozzle was set to 5\%, with a ratio of the turbulent-to-laminar viscosity of 100. To examine the influence of boundary layer transition location on the core flow parameters, simulations with different transition locations downstream of the nozzle throat were performed. The boundary layer transition location was varied until simulations with three different thermochemical models reached agreement with the experimental Pitot pressure data that was within experimental uncertainty of approximately 3\%. The final transition location is 380\,mm downstream of the nozzle throat. More information about the experimental facility and the process of obtaining Pitot pressure data can be found in \cite{Collen2021}.

\begin{table}[hbt!]
	\caption{\label{tab:stagnation-conditions} Stagnation conditions for the nozzle simulation. }
	\centering
	\begin{tabular}{ccc}
			\hline
			\hline
			Stagnation property     &  Value       & Unit \\\hline
		        \textit{p}$_{st}$          &  2.5469e+07  & Pa \\
		       \textit{ T}$_{st}$          &  2389.63     & K  \\  
     mass fraction [N$_2$]    &  0.76222     & -   \\  
     mass fraction [O$_2$]    &  0.22734     & -   \\  
     mass fraction [NO]       &  0.01041     & -   \\  
     mass fraction [O]        &  3.1125e-05  & -   \\  
     mass fraction [N]        &  0.0         & -   \\  
		    \hline
			\hline
		\end{tabular}
\end{table}

\begin{figure}[htbp!]
	\centering 
	\subfloat[]{\includegraphics[trim=1cm 0 0 0,clip,width=.56\textwidth]{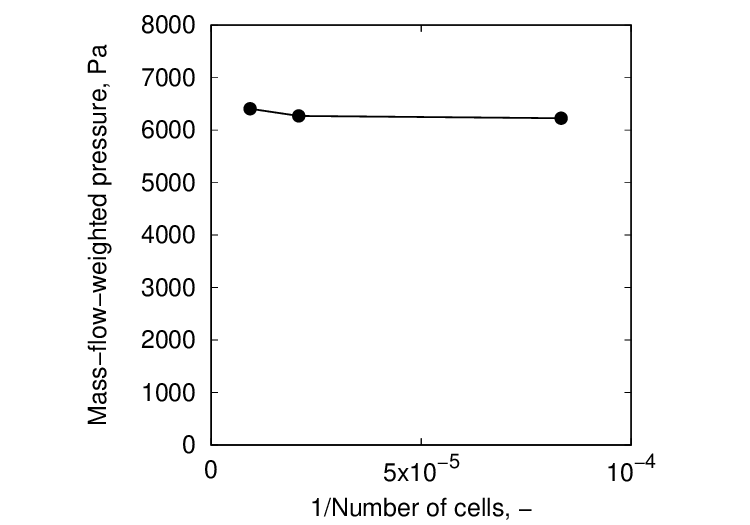}\label{fig:nozzle-grid-convergence}}
	\hspace{-60pt}
	\subfloat[]{\includegraphics[trim=0 0 1cm 0,clip,width=.56\textwidth]{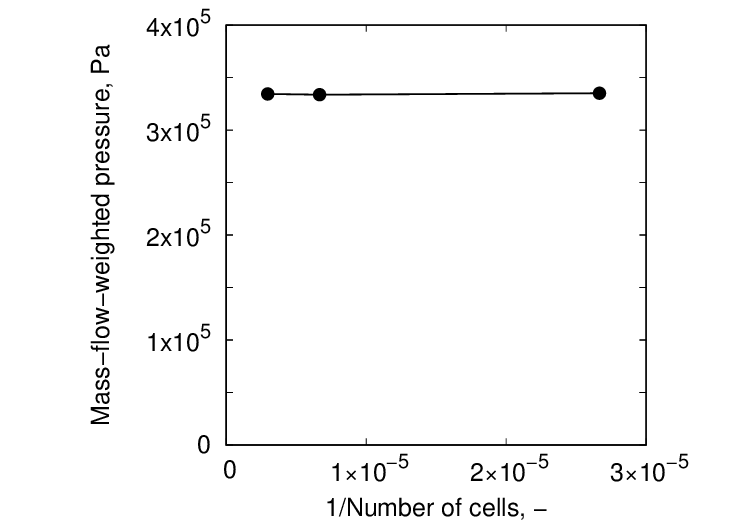}\label{fig:Pitot-grid-convergence}}
	\caption{Grid convergence analysis based on pressure for a) nozzle and b) Pitot probe simulations.}
	\label{fig:grid-convergence}
\end{figure}

\begin{figure}[htbp!]
	\centering
	\subfloat[]{\includegraphics[trim=0.0cm 0.0 0.0cm 0.35cm,clip,width=.33\textwidth]{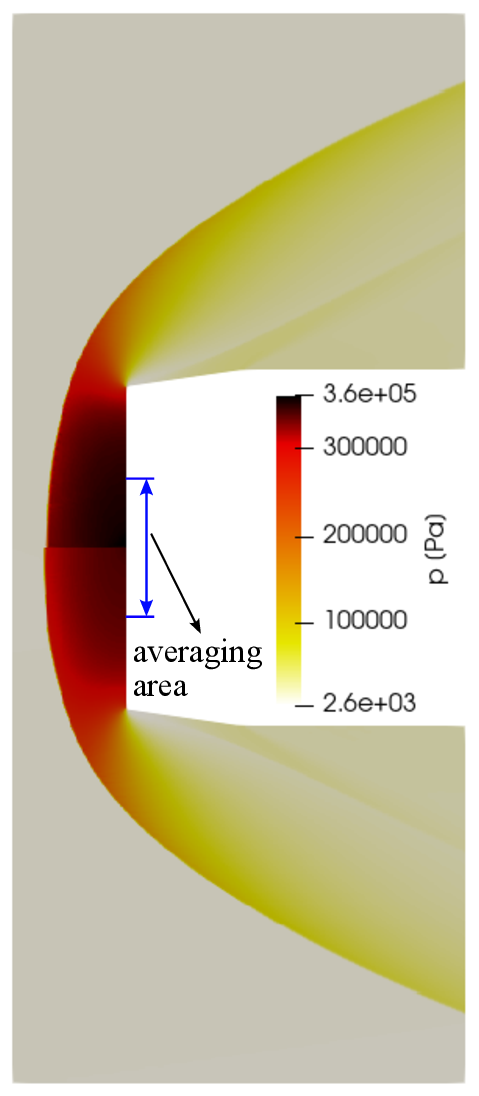}\label{fig:eq-noneq-comp-p}}
	\hfill
	\subfloat[]{\includegraphics[trim=0cm 0 0cm 0.45cm,clip,width=.33\textwidth]{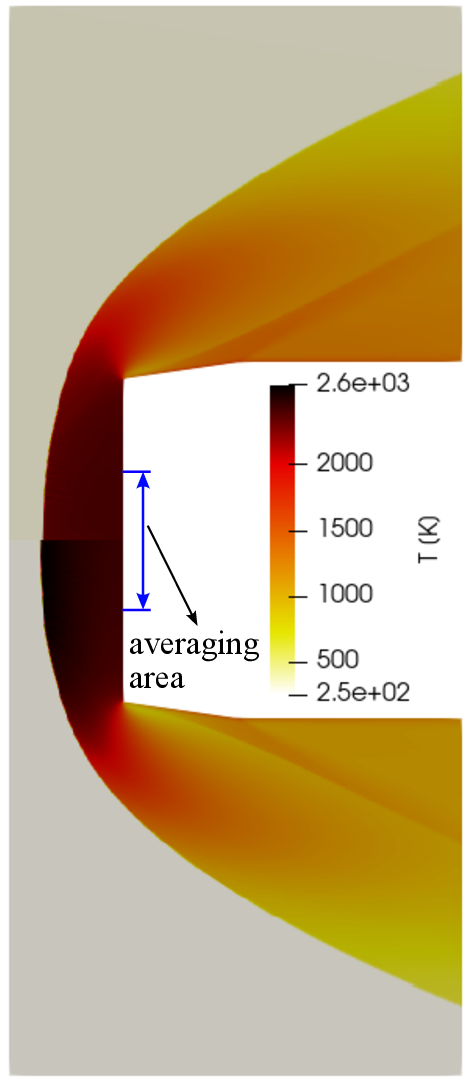}\label{fig:-eq-noneq-comp-T}}
    \hfill
	\subfloat[]{\includegraphics[trim=0.0cm 0.1cm 0cm 0.0cm,clip,width=.33\textwidth]{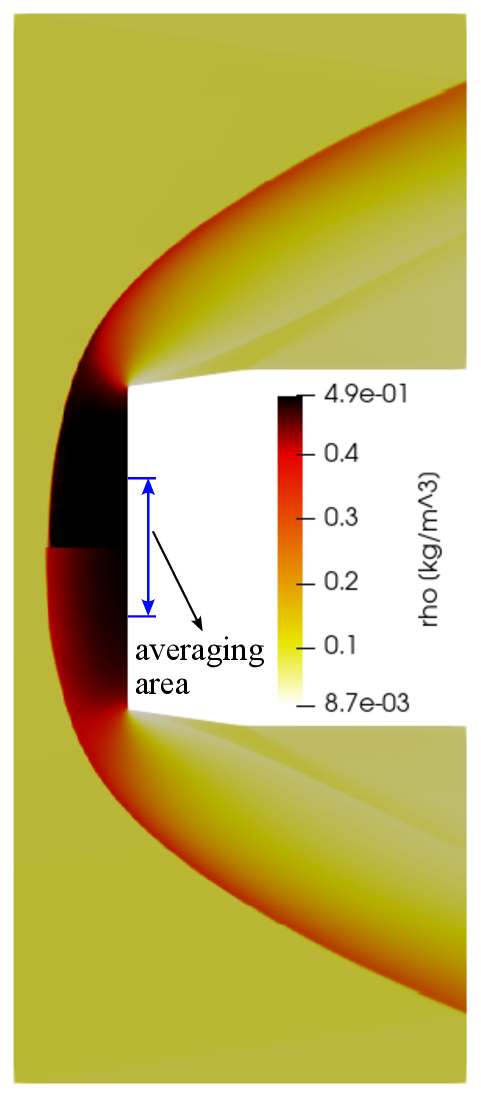}\label{fig:-eq-noneq-comp-rho}}
	\caption{Comparison of the equilibrium (top half) and the thermochemical non-equilibrium (bottom half) flow over Pitot probe. Flow variables are a) pressure, b) translational-rotational temperature, and c) density.}
	\label{fig:eq-noneq-comp}
\end{figure}

The computational grid which produced converged solutions for the flow simulations had 600 cells in the axial direction and 80 cells in the radial direction as shown in Fig.\ref{fig:nozzle-grid}. Grid clustering was employed near the throat and initial expansion regions to resolve the larger flow gradients expected in these regions and near the nozzle wall to ensure adequate resolution of the boundary layer. The grids near the nozzle wall were configured to be orthogonal to the wall to ensure that the near-wall flow parameters were properly approximated by the flow solver. The level of grid convergence is shown in Fig.\ref{fig:nozzle-grid-convergence} with a plot of mass-flow-weighted static pressure vs. the inverse of total number of cells for each grid resolution. The 
mass-flow-weighted static pressure value was obtained at the nozzle exit plane, with averaging area corresponding to the fraction of nozzle radius that would be equal to the radius of Pitot probe. The coarse grid had 300 axial cells and 40 radial cells, the medium grid had 600 axial cells and 80 radial cells, whilst the fine grid had 900 axial cells and 120 radial cells. The mass-flow-weighted pressure for the medium grid differed from the fine grid by only 0.7\%. This indicated that the medium grid produced a flow field at the exit of the nozzle that was sufficiently grid-converged for the analysis conducted in the present study at a cost of 1.56 hrs compute per flow solution. The output from the nozzle simulation was further used as a freestream inflow for a separate simulation of the Pitot probe, described in Sec~\ref{s:pitot-probe}. The position of the Pitot probe relative to the nozzle is shown in Fig.~\ref{fig:nozzle-and-probe}.

\subsection{Pitot probe}\label{s:pitot-probe}
The Pitot probe was modeled using a simplified geometry, as illustrated in Fig.~\ref{fig:Pitot-grid}. This did not account for the internal geometry which can lead to flow measurement fluctuations \cite{McGilvray2009}. However, the internal geometry of the Pitot probe is so small that it can be approximated that the sensor is near the surface of the probe. The probe was placed at an axial distance of 0.0033~m from the nozzle exit plane, approximating the position that would be used in the measurements. The nozzle and the probe have matching axes of symmetry (i.e. $y$~=~0). The Pitot probe near the stagnation point was modeled as inviscid flow because the flow is stagnated. There is no radius on the physical Pitot probe tip, or at worst, the radius is negligibly small, thus the probe was modeled with sharp edges. Grid clustering was employed around the wall. The grid independence was checked by plotting mass-flow weighted static pressure at the probe face against the inverse of the total number of cells for each grid resolution, as shown in Fig.~\ref{fig:Pitot-grid-convergence}. The mass-flow-weighted static pressure value was obtained at the probe face, with averaging area corresponding to the fraction of the probe face radius that would experience the flow. The coarse grid had the total of 37500, the medium grid 150000, and the fine grid 337500 cells. It can be seen in Fig.~\ref{fig:Pitot-grid-convergence} that the grid convergence was achieved as the difference in pressure levels for all three grid sizes is <1\%. The results reported here were obtained using the medium grid. The area with 1.5\,mm radius from the axis of symmetry is used for Pitot pressure averaging as this area is representative of the Pitot probe opening.

Figure~\ref{fig:eq-noneq-comp} shows a comparison of the flowfield for the equilibrium (top half of the image) and thermochemical non-equilibrium (bottom half of the image) flow over the Pitot probe, to illustrate the effect thermochemical non-equilibrium has on the freestream properties. Only equilibrium and thermochemical non-equilibrium are shown here as the differences between these two flowfields are the largest. Variables shown are pressure, temperature and density, while it is also possible to observe the difference in the bow shock stand-off distance for different thermochemistry. The largest difference in the flowfield values is observed in the region between the Pitot probe face and the bow shock. In this region, pressure plot demonstrates a slightly lower pressure for the thermochemical non-equilibrium compared to the equilibrium flow. Temperature is higher for the case of thermochemical non-equilibrium, while density is lower. As these flow properties give most of the information about the flowfield, it is clear from Fig.~\ref{fig:eq-noneq-comp} that it is necessary to consider the effect of thermochemical non-equilibrium on the flow.

\section{Pitot pressure predictions}
\begin{figure}[htbp!]\captionsetup[subfloat]{font=small,labelfont={bf},textfont={bf}}
	\centering 
	\subfloat[Numerical Pitot pressure obtained with coefficient C obtained from respective simulations using Rayleigh-Pitot equation and computed flow variables: C$_{\bf{EQ}}$=0.94388, C$_{\bf{CHEM.~NON-EQ}}$=0.94137, C$_{\bf{THERMOCHEM.~NON-EQ}}$=0.93002]{\includegraphics[trim=0 0 0 0,clip,width=0.7\textwidth]{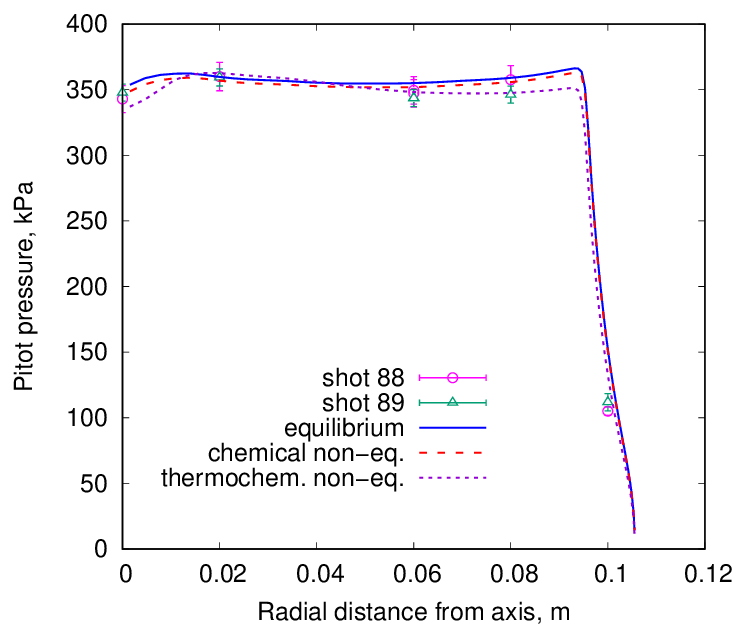}\label{fig:Pitot-pressure-profiles-real-C}}
\qquad
	\subfloat[Numerical Pitot pressure obtained with coefficient C equal to 0.92 for all cases.]{\includegraphics[trim=0 0 0 0,clip,width=0.7\textwidth]{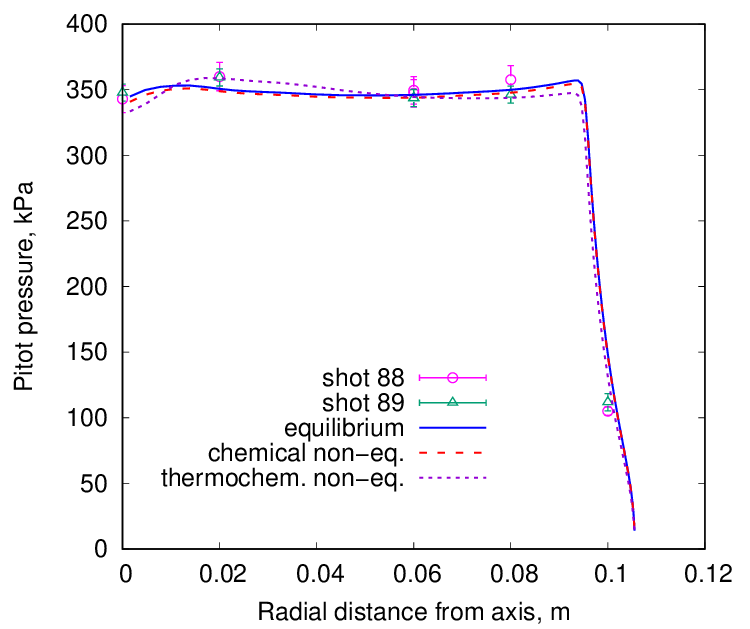}
	\label{fig:Pitot-pressure-profiles-092-C}}
\caption{Pitot pressure comparison between experimental and numerical data. Numerical results for all three cases were obtained for a boundary layer transition location set at 0.38\,m downstream of the nozzle throat.}
\label{fig:Pitot-pressure-profiles}
\end{figure}

 A comparison of experimental versus simulated Pitot pressure at the measurement location is provided in Fig.~\ref{fig:Pitot-pressure-profiles}. Numerical data are computed using the reduced form of the Rayleigh-Pitot equation and flow variables from the simulations with values taken at the centerline of the nozzle exit plane. The experimental values shown are measured Pitot pressure averaged over the test time. The results show good uniformity across the core flow region, which the experimental data suggests has a diameter of at least 160\,mm. In Fig.~\ref{fig:Pitot-pressure-profiles-real-C}, the numerical $\textit{p}_{Pitot}$ values are computed via the reduced form of the Rayleigh-Pitot formula and using the C values obtained from the respective simulations at the centerline of the nozzle exit plane. Fig.~\ref{fig:Pitot-pressure-profiles-092-C} shows numerical Pitot pressures obtained using Eq.~\ref{eq:Pitot} with a C value of 0.92 for all three thermochemical models to illustrate the difference when this value is applied to all cases, compared to Fig.~\ref{fig:Pitot-pressure-profiles-real-C} where appropriately tuned C is applied. Fig.~\ref{fig:Pitot-pressure-profiles-092-C} was previously reported in Ref.\cite{Collen2021}, an overview of the T6 Stalker facility. Table~\ref{tab:constants-obtained} provides C values for all three thermochemical models obtained from Pitot probe simulations, where tabulated C values were obtained from the post-shock stagnation point conditions of the Pitot probe simulation using a respective thermochemical model. These computed C values show a dependence on the non-equilibrium relaxation effects such that C values, and thus computed Pitot pressures, decrease with the increase in relaxation times, a result that is consistent with the findings of Ref.~\cite{Kantrowitz1946,Huber1947}.

Fig.~\ref{fig:Pitot-pressure-profiles} demonstrates that overall, there is an excellent match between the experiments and the simulations. However, Fig.~\ref{fig:Pitot-pressure-profiles-real-C} illustrates that within the core flow region, the thermochemical non-equilibrium result shows slightly better agreement with measurement data points. Additionally, it is shown that by using the tuned coefficient values to compute Pitot pressure, the resulting values for Pitot pressures for all three thermochemical models agree somewhat better with measurement data points than when using C~=~0.92 for all cases. While these improvements in agreement might be considered small judging from the total values only, they become more significant considering that the experimental errors (as reported on the measurement data points) are already quite low. It will be shown in the following sections what is the effect on the total test flow properties.

Most experimenters calculate Pitot pressures using the Rayleigh-Pitot relation (Eq.~\ref{eq:Rayleigh-Pitot}) or its reduced form (Eq.~\ref{eq:Pitot}) for ideal air at high Mach number (with the assumption of a constant ratio of specific heats of 1.4). However, applying the Rayleigh-Pitot relation (Eq.~\ref{eq:Rayleigh-Pitot}) in the thermochemical non-equilibrium case resulted in Pitot pressure values that were not in agreement with the other two thermochemistry models, which is not surprising considering that the Rayleigh-Pitot equation was derived for an ideal gas (both thermally and calorically perfect). 

\begin{table}[hbt!]
	\caption{\label{tab:constants-obtained} Constant in Rayleigh-Pitot equation for high Mach numbers, obtained for different thermochemical calculations for both nozzle and Pitot probe. }
	\centering
	\begin{tabular}{cc}
			\hline
			\hline
		        Thermochemical model       & {Constant C} \\\hline
		              equilibrium          & 0.94388  \\
		       chemical non-equilibrium    & 0.94137 \\ 
		   thermochemical non-equilibrium  & 0.93002 \\ 
		    \hline
			\hline
		\end{tabular}
\end{table}

Table~\ref{tab:area-avg-pitot} presents Pitot pressure values obtained from simulations with different combinations of thermochemistry models in the nozzle simulations and the subsequent Pitot probe simulations, i.e., the thermochemistry model in nozzle simulation for some cases differed from the thermochemistry model of the Pitot probe simulation. This resulted in a maximum of 5\% difference on the Pitot pressures (as shown in the matching Pitot pressure levels in Fig.~\ref{fig:Pitot-pressure-profiles}). This is not surprising since the Pitot pressure value is driven by the momentum of the gas which is not greatly influenced by the detailed distribution of energy within the gas stream. 
\setlength{\tabcolsep}{3.5pt}
\begin{table}[hbt!]
	\caption{\label{tab:area-avg-pitot} Pitot pressure averaged over area of the Pitot probe opening of $\phi$=3\,mm. Expressed in percentage of difference off the nominal thermochemical non-equilibrium value of 337\,kPa.}
	\vspace{-5pt}
	\centering
	\begin{tabular}{ llcccc }
			\toprule
			\toprule
		    \multicolumn{2}{l}{} & 0.92$\rho U^2$ & EQ & CHEM. NON-EQ & THERMOCHEM. NON-EQ \\ 
		    \cmidrule(r){2-6}

\multirow{3}{*}{\rotatebox[origin=c]{90}{\parbox[c]{1cm}{\centering \bf{nozzle sim.}}}}   
&          EQ\textsc{uilibrium}        &  2.28\% & 4.96\% &  4.12\% & 3.29\% \\ 
& CHEM. NON-EQ\textsc{uilibrium}       &  1.09\% & 3.38\% &  3.46\% & 2.81\% \\
& THERMOCHEM. NON-EQ\textsc{uilibrium} & -1.09\% & 0.66\% &  0.87\%  & 0.00\%  \\
			\hline
			\hline
		\end{tabular}
	\end{table}

\section{Effect on freestream conditions}
 
While different thermochemistry appears to have little effect on Pitot pressure, it is expected to have a more significant effect on other flow properties. To explain this, simulations are undertaken for the thermochemical non-equilibrium case. In this work, we are assuming the correct value of the Pitot pressure is the one obtained through the same chemistry assumption for both nozzle and Pitot probe. The correct method is to tune to the Pitot probe measurement to obtain correct freestream values. This is done by changing the boundary layer transition location in the nozzle. Again, we are following the typical approach of an experimenter where reduced form of the Rayleigh-Pitot equation with 0.92 is used to compute numerical Pitot pressure that we are trying to match to the measured Pitot pressure. 
 
The results of this process are listed in Table~\ref{tab:freestream-discrepancies-noneq}, where the baseline case is the thermochemical non-equilibrium nozzle and Pitot probe, resulting in Pitot pressure of 337\,kPa. The percentage differences in columns 2 and 3 in Table~\ref{tab:freestream-discrepancies-noneq} were calculated using the following equation:

\begin{equation}
	\Delta V_{ESTCN} [\%] = \frac{V_{ESTCN}-V_{NON-EQ}}{ V_{NON-EQ} } \cdot 100
	\label{difference}	
\end{equation}

where $V$ is a freestream property, such as density, velocity, static pressure, etc. Subscript ESTCN stands for calculations performed using ESTCN with either non-equilibrium or equilibrium gas model, while subscript NON-EQ relates to the result of the non-equilibrium nozzle and Pitot probe simulations (i.e., the baseline case).

The difference in freestream values (column 3) is not massive, however, the tuned transition location is only a few millimetres away from the initial one, and so the values would not be significantly different between the initial (column 2) and the transition-location-tuned simulation (column 3). The reason for this is that the transition location was already tuned to match experimental Pitot pressure, as explained in Sec.~\ref{s:pitot-probe} and shown in Fig.~\ref{fig:Pitot-pressure-profiles}. Thus, the process explained here was the fine tuning of the transition location and Pitot pressure. The resulting transition location is only 3\,mm upstream from the previous location of 380\,mm, i.e., the resulting transition location is now 377\,mm. If less effort was put initially in tuning the transition location, then naturally the process in this section would be longer and the obtained differences in freestream properties would be larger.

\newcolumntype{d}[1]{D..{#1}} 
\setlength{\tabcolsep}{15pt}
\begin{table}[hbt!]
	\caption{\label{tab:freestream-discrepancies-noneq} Discrepancies in freestream values for the non-equilibrium  and equilibrium nozzle. }
	\vspace{-5pt}
	\centering
	\begin{tabular}{ lccc }
			\toprule
			\toprule	
			\multirow{2}*{Property} & \multirow{2}*{\text{NON-EQ}} &  \text{$\Delta$NON-EQ nozzle}  &  \text{$\Delta$EQ nozzle}\\ 
			& &\text{($p_{Pitot}$=0.92$\rho U^2$)} &\text{($p_{Pitot}$=0.92$\rho U^2$)} \\
			\cmidrule(){1-4}
			$\rho$ [kg/m$^3$] & 0.074 &  0.8\%    & -2.2\%\\
			$U$ [m/s]         & 2214  &  -0.02\%  &  1.6\%\\
			$p$ [kPa]         & 5390  &  1.1\%    & 11.9\%\\   
			$T_{tr}$ [K]      & 253   &  0.3\%    & 14.4\%\\   
			$T_{ve}$ [K]      & 1271  &   0.01\%  & ---\\     
			$M$    [-]        & 7.1   &  -0.2\%   & -7.7\%\\   
			\hline
			\hline
		\end{tabular}
	\end{table}

Table~\ref{tab:freestream-discrepancies-noneq} also includes equilibrium nozzle calculations. As the experimenters typically assume equilibrium flow properties at low temperatures, the freestream conditions are generally assessed assuming the measured Pitot pressure and tuning an equilibrium isentropic nozzle calculation to match that value. Experimenters in reflected-shock tunnels usually perform this equilibrium nozzle calculation using a state-to-state calculation program, such as ESTCN \cite{Jacobs2011}, for estimating flow conditions. ESTCN simulations were run targeting the Pitot pressure of the baseline case (non-equilibrium thermochemistry simulations). Pitot pressure was computed using the reduced form of the Rayleigh-Pitot equation with constant C equal to 0.92. It is evident from the values in Table~\ref{tab:freestream-discrepancies-noneq} that using the equilibrium calculations with C~=~0.92 to obtain Pitot pressure results in higher freestream discrepancies compared to the baseline case than when the non-equilibrium nozzle simulations with tuning the transition location are performed. 

A sweep of free-stream conditions was performed, varying only enthalpy and using the same Mach 7 nozzle. These calculations were done for air using the following approach: first, one-dimensional nozzle simulations were done using NENZF1d \footnote{NENZF1d code is a part of the Gas Dynamics Toolkit and freely available at \url{https://gdtk.uqcloud.net}}. Second, using the nozzle simulation result as an inflow, axisymmetric simulations were performed for a sphere, and third, C values were obtained directly from the observed simulated pressure in the cells against the stagnation point on the sphere as in the Eq.~\ref{eq:sphere}: 

 \begin{equation}
 \label{eq:sphere}
  C = \frac{\textit{p}_{simulation}}{\rho \textit{U}^2},
\end{equation}

where $\rho$ and $\textit{U}$ are freestream conditions used in the simulation.

These simulations were done for both one-temperature thermal model with finite-rate chemistry (based on Gupta's 5-species air model \cite{Gupta1990}) and two-temperature thermal-non-equilibrium model with finite-rate chemistry (based on Park's 2-temperature 5-species air model \cite{Gupta1990}). The axisymmetric computational grid was formed with 2400 cells.

\begin{figure}[H]
	\centering
	\includegraphics[width=0.97\textwidth]{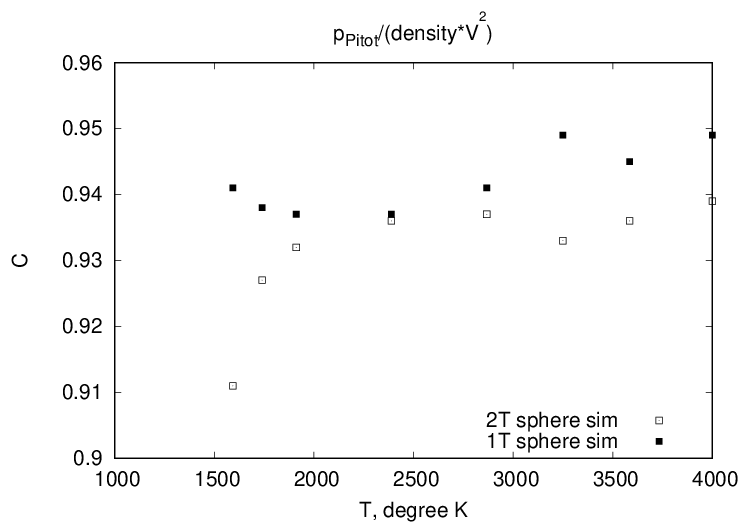} 
	\caption{Variation of C coefficient for a range of test conditions.}
	\label{fig:enthalpy-vs-C}
\end{figure}
\vspace{-3mm}
The results are shown in Fig.~\ref{fig:enthalpy-vs-C} and illustrate the variation of coefficient C over a range of test conditions (i.e., enthalpies) and for two different thermochemical models. It is demonstrated that the variation of the coefficient C is most significant at lower enthalpies up to approximately 2.6~MJ/kg for the two-temperature thermal-non-equilibrium model, after which it asymptotes at a value C~$\approx$~0.94. The data for one-temperature thermal model is approximately 0.94. While there seems to be no reliable trend for C vs. enthalpy, the approximate level of about 0.94 presents a useful starting data point for researchers which can then be refined for any particular operating point using a single blunt-body calculation. The commonly used value C~=~0.92 is not correct for all conditions, particularly at enthalpies higher than approximately 2.6~MJ/kg, which makes up for the significant amount of tests done in shock tunnels. 

\section{Conclusion}

In this study, a range of CFD simulations were conducted to simulate flow in a hypersonic nozzle. The nozzle-exit flow properties from these simulations displayed departure from the equilibrium, highlighting the need to take the influence of finite-rate thermochemistry into account. These results show that the effects of thermochemical non-equilibrium are much stronger than expected at a relatively low enthalpy condition of 2.67~MJ/kg. The values of constant C in the reduced formulation of the Rayleigh-Pitot equation computed from our simulations show departure from the commonly used value of 0.92. These computed C values show
a dependence on the non-equilibrium relaxation effects such that C values, and thus computed Pitot pressures, decrease with the increase in relaxation times, a result that is consistent with the findings of Ref.~\cite{Kantrowitz1946,Huber1947}. 

Our results show that the calculation of Pitot pressures using 0.92 value for C lead to a maximum of 5\% difference, a result that is not surprising since values of Pitot pressure are driven by the momentum of the gas which is not greatly influenced by the detailed distribution of energy within the gas stream. However, our results also show that equilibrium calculations using C~=~0.92 lead to much higher discrepancies in computed freestream properties compared to the case than when the non-equilibrium nozzle simulations with tuning the transition location are performed. The most significant effect is on the static pressure and temperature with 11.9\% and 14.4\% difference, respectively. This aligns with results from other reflected shock tunnel facilities at similar enthalpies, for example \cite{Hannemann2010a}. Future experiments are planned to confirm the degree of thermal non-equilibrium in the flow field, both through optical techniques and measurement of the freestream static pressure.

Additional calculations were done with a sweep of free-stream conditions, varying only enthalpy. While the results did not show a reliable trend for C vs. enthalpy, the approximate level of about 0.94 emerged from the data. This value of C presents a useful starting data point for researchers and can be further refined for any particular operating point using a single blunt-body calculation. We recommend that suitable values for C should be estimated for each operation point of a shock tunnel facility.

\section*{Appendix}
\label{Appendix}
 Across a normal shock wave, such as a bow shock at the nose of an aircraft, maximum pressure coefficient is:
\begin{equation}
	C_{p,\,max} =\frac{\textit{p}_{0,2}-\textit{p}_{\infty}}{\frac{1}{2} \rho_{\infty} \textit{U}_{\infty}^2}  
	\label{Cpmax}
\end{equation}

\noindent Rewriting this in terms of freestream Mach number for supersonic conditions:
\begin{equation}
	C_{p,\,max} = \frac{\textit{p}_{0,2}-\textit{p}_{\infty}}{\frac{1}{2} \rho_{\infty} \textit{U}_{\infty}^2} 
	= \frac{\textit{p}_{0,2}-\textit{p}_{\infty}}{\frac{\gamma}{2} \textit{p}_{\infty} M_{\infty}^2} 
	= \frac{1}{{\frac{\gamma}{2}} M_{\infty}^2} 
	\left[ \frac{ \left(\frac{\gamma + 1}{2} M_{\infty}^2\right) ^{\left(\frac{\gamma}{\gamma -1}\right)}}
	{\left(\frac{2 \gamma}{\gamma +1} M_{\infty}^2 - \frac{\gamma -1}{\gamma +1} \right)^{\frac{1}{\gamma -1}} } -1
	\right]
	\label{Cpmax-supersonic}	
\end{equation}

\noindent For an ``infinite'' Mach number this becomes:
\allowdisplaybreaks
\begin{equation}
	\begin{split}
		\frac{C_{p,\,max}}{\lim\limits_{M_{\infty}\to\infty}} &= \frac{1}{{\frac{\gamma}{2}} \left(\frac{M_{\infty}}{\lim\limits_{M_{\infty}\to\infty}} \right)^2} 
		\left[ \frac{ \left(\frac{\gamma + 1}{2}
			\left(\frac{M_{\infty}}{\lim\limits_{M_{\infty}\to\infty}} \right)^2 \right) ^{\left(\frac{\gamma}{\gamma -1}\right)}}
		{\left(\frac{2 \gamma}{\gamma +1} 
			\left(\frac{M_{\infty}}{\lim\limits_{M_{\infty}\to\infty}} \right)^2 - \frac{\gamma -1}{\gamma +1} \right)^{\frac{1}{\gamma -1}} } -1 \right] \\
		&= \frac{1}{{\frac{\gamma}{2}} \left(\frac{M_{\infty}}{\lim\limits_{M_{\infty}\to\infty}} \right)^2} 
		\left[ \frac{ \left(\frac{\gamma + 1}{2}
			\left(\frac{M_{\infty}}{\lim\limits_{M_{\infty}\to\infty}} \right)^2 \right) ^{\left(\frac{\gamma}{\gamma -1}\right)}}
		{\left(\frac{2 \gamma}{\gamma +1} 
			\left(\frac{M_{\infty}}{\lim\limits_{M_{\infty}\to\infty}} \right)^2 \right)^{\left(\frac{1}{\gamma -1}\right)} } \right] \\
		&= \frac{\frac{2}{\gamma} {\left(\frac{\gamma + 1}{2} \right)}^{\left(\frac{\gamma}{\gamma - 1} \right)} }{{\left( \frac{2 \gamma}{\gamma + 1}\right)} ^ {\left( \frac{1}{\gamma -1} \right)}} 
		\left[ \frac{ 
			\left(\frac{M_{\infty}}{\lim\limits_{M_{\infty}\to\infty}} \right) ^{\left(\frac{2 \gamma}{\gamma -1}\right)}}
		{\left( \left(\frac{M_{\infty}}{\lim\limits_{M_{\infty}\to\infty}} \right) ^{2(\gamma -1)}
			\left(\frac{M_{\infty}}{\lim\limits_{M_{\infty}\to\infty}} \right)^2 \right)^{\frac{1}{\gamma -1}} } \right]
		\label{Cpmax-infinite-Mach}
	\end{split}
\end{equation}

\noindent Collecting exponents on  \begin{math} \frac{M_{\infty}}{\lim\limits_{M_{\infty}\to\infty}} \end{math}: 
\begin{equation}
	\begin{split}
		\frac{C_{p,\,max}}{\lim\limits_{M_{\infty}\to\infty}} &= \frac{\frac{2}{\gamma} {\left(\frac{\gamma + 1}{2} \right)}^{\left(\frac{\gamma}{\gamma -1} \right)} }{{\left(\frac{2 \gamma}{\gamma +1}\right)} 
			^ {\left( \frac{1}{\gamma -1} \right)}} 
		\left[ \frac{M_{\infty}}{\lim\limits_{M_{\infty}\to\infty}}
		^ {\left(\frac{2 \gamma}{\gamma -1} \right)
			-(2(\gamma -1)+2)
			\left(\frac{1}{\gamma -1} \right)} 		\right] \\
		&= \frac{\frac{2}{\gamma} {\left( \frac{\gamma +1}{2} \right)} ^ {\left( \frac{\gamma}{\gamma -1} \right)} }
		{{\left(\frac{2 \gamma}{\gamma +1} \right)} ^ {\left( \frac{1}{\gamma -1} \right)}} 
		\left[ \left( \frac{ M_{\infty} }{ \lim\limits_{M_{\infty}\to\infty} } \right) ^ 0 \right] \\
		&= \frac{\frac{2}{\gamma} {\left( \frac{\gamma +1}{2} \right)} ^ {\left( \frac{\gamma}{\gamma -1} \right)} }
		{{\left(\frac{2 \gamma}{\gamma +1} \right)} ^ {\left( \frac{1}{\gamma -1} \right)}}  \\
		\label{Cpmax-infinite-Mach-2}
	\end{split}
\end{equation}

\noindent Then doing some manipulations: 
\begin{equation}
	\begin{split}
		\frac{C_{p,\,max}}{\lim\limits_{M_{\infty}\to\infty}} &= \frac{1}{\gamma} \cdot 
		\frac{2}{2^{\left( \frac{\gamma}{\gamma -1} \right)} \cdot 2^{\left( \frac{1}{\gamma -1} \right)}} \cdot
		\frac{{(\gamma +1) ^ {\left(\frac{\gamma}{\gamma -1} \right)} }}{{\left( \frac{\gamma}{\gamma +1} \right)} ^ {\left( \frac{1}{\gamma -1} \right)} } \\
		&= \frac{1}{\gamma} \cdot 
		\frac{1}{2 ^ {\left( \frac{2}{\gamma -1} \right) } } \cdot 
		\frac{ (\gamma +1) ^ {\left(\frac{\gamma}{\gamma -1}\right) } }{ {\left( \frac{\gamma}{\gamma +1} \right)} ^ {\left( \frac{1}{\gamma -1} \right)} } \\
		&= \frac{1}{\gamma ^ {\left( 1+ \frac{1}{\gamma -1} \right) }} \cdot
		\frac{1}{2 ^ {\left( \frac{2}{\gamma -1} \right) }} 
		(\gamma +1) ^ {\left( {\frac{\gamma}{\gamma -1} } + \gamma -1 \right)} \\
		&= \frac{\gamma ^{\left( \frac{\gamma}{1- \gamma} \right)} }{2^ {\left( \frac{2}{\gamma -1} \right)} } 
		(\gamma +1) ^ {\left( \frac{\gamma +1}{\gamma -1} \right)}
		\label{Cpmax-infinite-Mach-3}
	\end{split}
\end{equation}

\noindent If we substitute $\gamma$=1.4 in Eq.\ref{Cpmax-infinite-Mach-3} then $C_{p\,max}$ becomes:
\begin{equation}
	\begin{split}
		\frac{C_{p,\,max}}{\lim\limits_{M_{\infty}\to\infty}} = 1.8394
		\label{Cpmax-infinite-Mach-4}
	\end{split}
\end{equation}

\bibliography{bibliography}

\end{document}